\documentclass[prd, reprint, preprintnumbers, twocolumn, eqsecnum,floatfix,letterpaper,superscriptaddress,nofootinbib]{revtex4}
\usepackage{mathtools}
\usepackage{amsmath,amssymb,graphicx}
\usepackage{xcolor}
\usepackage{graphicx}

\graphicspath{{./}{figures/}}
\usepackage[colorlinks, pdfborder={0 0 0}]{hyperref}

\definecolor{LinkColor}{rgb}{0.75 , 0, 0}
\definecolor{CiteColor}{rgb}{0, 0.5, 0.5}
\definecolor{UrlColor}{rgb}{0, 0, 0.75}
\hypersetup{linkcolor=LinkColor}
\hypersetup{citecolor=CiteColor}
\hypersetup{urlcolor=UrlColor}

\begin{document}

\title{Demonstration of Machine Learning-assisted real-time noise regression in gravitational wave detectors}
\author{Muhammed Saleem}\email{mcholayi@umn.edu} 
\affiliation{School of Physics and Astronomy, University of Minnesota, Minneapolis, MN 55455, USA}
\author{Alec Gunny}
\affiliation{Department of Physics, MIT, Cambridge, MA 02139, USA}
\affiliation{LIGO Laboratory, 185 Albany St, MIT, Cambridge, MA 02139, USA}
\author{Chia-Jui Chou}
\affiliation{Department of Electrophysics, National Yang Ming Chiao Tung University, Hsinchu, Taiwan}
\author{Li-Cheng Yang}
\affiliation{Department of Electrophysics, National Yang Ming Chiao Tung University, Hsinchu, Taiwan}
\author{Shu-Wei Yeh}
\affiliation{Department of Physics, National Tsing Hua University, Hsinchu, Taiwan}
\author{Andy H. Y. Chen}
\affiliation{Institute of Physics, National Yang Ming Chiao Tung University, Hsinchu, Taiwan}
\author{Ryan Magee}
\affiliation{LIGO Laboratory, California Institute of Technology, Pasadena, CA 91125, USA}
\author{William Benoit}
\affiliation{School of Physics and Astronomy, University of Minnesota, Minneapolis, MN 55455, USA}
\author{Tri Nguyen}
\affiliation{Department of Physics, MIT, Cambridge, MA 02139, USA}
\affiliation{LIGO Laboratory, 185 Albany St, MIT, Cambridge, MA 02139, USA}
\author{Pinchen Fan}
\affiliation{Department of Physics, MIT, Cambridge, MA 02139, USA}
\affiliation{LIGO Laboratory, 185 Albany St, MIT, Cambridge, MA 02139, USA}
\author{Deep Chatterjee}
\affiliation{Department of Physics, MIT, Cambridge, MA 02139, USA}
\affiliation{LIGO Laboratory, 185 Albany St, MIT, Cambridge, MA 02139, USA}
\author{Ethan Marx}
\affiliation{Department of Physics, MIT, Cambridge, MA 02139, USA}
\affiliation{LIGO Laboratory, 185 Albany St, MIT, Cambridge, MA 02139, USA}
\author{Eric Moreno}
\affiliation{Department of Physics, MIT, Cambridge, MA 02139, USA}
\affiliation{LIGO Laboratory, 185 Albany St, MIT, Cambridge, MA 02139, USA}
\author{Rafia Omer}
\affiliation{School of Physics and Astronomy, University of Minnesota, Minneapolis, MN 55455, USA}
\author{Ryan Raikman}
\affiliation{Department of Physics, MIT, Cambridge, MA 02139, USA}
\affiliation{LIGO Laboratory, 185 Albany St, MIT, Cambridge, MA 02139, USA}
\author{Dylan Rankin}
\affiliation{Department of Physics, MIT, Cambridge, MA 02139, USA}
\affiliation{LIGO Laboratory, 185 Albany St, MIT, Cambridge, MA 02139, USA}
\author{Ritwik Sharma}
\affiliation{Department of Physics,
Deshbandhu College, University of Delhi, New Delhi, India}
\author{Michael Coughlin}
\affiliation{School of Physics and Astronomy, University of Minnesota, Minneapolis, MN 55455, USA}
\author{Philip Harris}
\author{Erik Katsavounidis}
\affiliation{Department of Physics, MIT, Cambridge, MA 02139, USA}
\affiliation{LIGO Laboratory, 185 Albany St, MIT, Cambridge, MA 02139, USA}

\date{\today}


\begin{abstract}

Real-time noise regression algorithms are crucial for maximizing the science outcomes of the LIGO, Virgo, and KAGRA gravitational-wave detectors. This includes improvements in the detectability, source localization and pre-merger detectability of signals thereby enabling rapid multi-messenger follow-up. In this paper, we demonstrate the effectiveness of \textit{DeepClean}, a convolutional neural network architecture that uses witness sensors to estimate and subtract non-linear and non-stationary noise from gravitational-wave strain data. Our study uses LIGO data from the third observing run with injected compact binary signals. As a demonstration, we use \textit{DeepClean} to subtract the noise at 60 Hz due to the power mains and their sidebands arising from non-linear coupling with other instrumental noise sources. Our parameter estimation study on the injected signals shows that \textit{DeepClean} does not do any harm to the underlying astrophysical signals in the data while it can enhances the signal-to-noise ratio of potential signals. We show that  \textit{DeepClean} can be used for low-latency noise regression to produce cleaned output data at latencies $\sim 1-2$\, s. We also discuss various considerations that may be made while training  \textit{DeepClean} for low latency applications.  

\end{abstract}
 \maketitle

\section{Introduction} \label{sec:intro}

The current network of ground-based laser interferometers, consisting of advanced LIGO ~\cite{AdLIGO,ligoHarry_2010} and advanced Virgo ~\cite{AdVIRGO} have facilitated the detection of approximately one hundred gravitational wave (GW) events from coalescing compact binaries consisting of neutron stars and/or black holes~\citep{GWTC3,Nitz:2021zwj,Olsen:2022pin}. In the third observing run (referred to as O3), LIGO Livingston (L1), LIGO Hanford (H1)  and Virgo (V1) had a sensitive median range for detecting binary neutron stars (BNS) of approximately 133\,Mpc, 115\,Mpc and 51 \,Mpc, respectively~\citep{GWTC3}. The fourth observing run, referred to as O4, has been officially started in May 2023, with recent technological upgrades and the addition of a fourth detector, KAGRA (K1)~\citep{kagraPhysRevD.88.043007,Som2012}. The range of the anticipated detection of binary neutron stars in O4 is expected to be 160-190\,Mpc for Advanced LIGO, 90-120\,Mpc for Advanced Virgo, and 1-10\,Mpc for KAGRA~\citep{OSD}.

Upgrades in technology have improved the sensitivity of interferometers by reducing fundamental noise sources such as thermal and quantum fluctuations~\citep{Cahillane:2022pqm,DaAr2021}. However, environmental and instrumental processes also contribute to the noise in the interferometer strain. The presence of such noise can reduce the sensitivity of the detectors to astrophysical transient signals~\cite{LIGO:2021ppb,2021CQGra..38b5016S}, in particular, sources without well-known theoretical models (e.g., supernovae)~\citep{Riles_2013}. Noise regression methods are used to remove these contaminants, typically by identifying their origin~\cite{Tiwari:2015ofa}. Gravitational-wave interferometers are equipped with auxiliary \textit{witness} sensors or channels to independently monitor these processes in addition to the strain channel~\cite{2021CQGra..38n5001N}. Identifying the couplings that exist between these witness channels is the key in estimating their contribution to the strain and removing them. However, there are thousands of witness channels tracking different noise sources, and non-linear couplings between them may result in noise that is challenging to identify using standard filtering techniques such as Wiener filtering~\citep{Vas2001,Say2003,Davis:2018yrz,LIGOScientific:2019hgc}.

The developments of machine learning neural networks have significantly enhanced our capability of noise regression in the interferometer strain data. This includes the recent deep learning algorithms that are developed to subtract non-linear and non-stationary couplings originating from instrumental and environmental sources~\citep{BiBl2013,ZeCo2017,PhysRevD.95.104059,Vajente_2020,Ormiston:2020ele,Yu:2021swq}. These algorithms have successfully removed noise couplings such as the 60\,Hz power-line noise and their sidebands, which arise from the non-linear coupling of the strain with instrumental control systems. However, these deep learning noise regression algorithms have thus far been demonstrated primarily in high latency, or \textit{offline}, analysis scenarios, where time-series data of several hours are analyzed long after they were originally recorded.

Multi-messenger astronomy, where gravitational-wave sources are followed up for their counterparts in the electromagnetic spectrum and neutrinos, is one of the most promising aspects of gravitational-wave observations~\citep{Abbott:2016blz,AbEA2017b}. Detecting electromagnetic counterparts that fade quickly after the gravitational-wave detection, such as $\gamma$-ray bursts and x-rays from binary neutron star mergers, requires sending out low-latency alerts to trigger follow-up observations across electromagnetic frequencies~\citep{Metzger:2016pju,Cannon:2011vi,LIGOScientific:2019gag,Sachdev:2020lfd,Chu:2020pjv,Yu:2021vvm}. Ground-based detectors are still below their designed sensitivities at lower frequency ranges (below 60\,Hz)~\citep{Abbott:2016xvh,aLIGO:2020wna}, indicating the potential for substantial improvements in the capability of sending pre-merger (or early-warning) alerts by performing low-latency noise regression at low frequencies. Even incremental improvements in the sensitive distances can lead to significant improvements in the number of detections, which scale as the cube of the distance (proportional to the volume). These improvements could result in the detection of compact binary mergers that would not have otherwise been identified at low latency.

Performing low-latency  (\textit{a.k.a} online) noise regression poses significant computational challenges compared to offline regression. To not become the dominant source of latency in the release of alerts, a low-latency noise regression should produce cleaned strain with the overall delay not more than a couple of seconds. Ref.~\cite{Gunny:2021gne} discussed in detail how to meet  such computational demands in low latency, by employing the \textit{as-a-service} computing paradigm~\citep{Krupa:2020bwg,Wang:2020fjr} into the context of gravitational-wave data analysis, in order to leverage hardware accelerators (such as GPUs) and other heterogeneous computing resources.

In this paper, we demonstrate and validate the application of \textit{DeepClean}~\cite{Ormiston:2020ele} infrastructure on gravitational wave strain data from LIGO Hanford and LIGO Livingston. \textit{DeepClean} is a deep learning convolutional neural network algorithm for noise regression in gravitational-wave strain. \textit{DeepClean} targets those noise that are environmental or technical\footnote{Technical noise, \textit{a.k.a} control noise, usually refers to the noise generated by the apparatus that control the optics in the interferometer} in origin and can be tracked independently with witness sensors. We perform a mock data challenge (MDC) to demonstrate the effectiveness of \textit{DeepClean} as a production pipeline for low-latency and high-latency noise regression applications.

This paper is organized as follows:
Section~\ref{sec:infra} provides a concise overview of the DeepClean architecture and the end-to-end infrastructure.
Section~\ref{sec-MDC} presents the details of our mock data challenge.
In section~\ref{sec:inference}, we delve into the application of \textit{DeepClean} on our mock data and present the corresponding performance metrics.
Section~\ref{sec:offline-O3} demonstrates the validation tests performed using astrophysically motivated metrics, including detection and parameter estimation.
Section~\ref{sec:dc-online} focuses on the feasibility of utilizing \textit{DeepClean} for low-latency noise regression.
Finally, section~\ref{sec:conclusions} concludes the study and discusses future prospects.

\begin{figure*}
    \centering
    \includegraphics[scale=0.36]{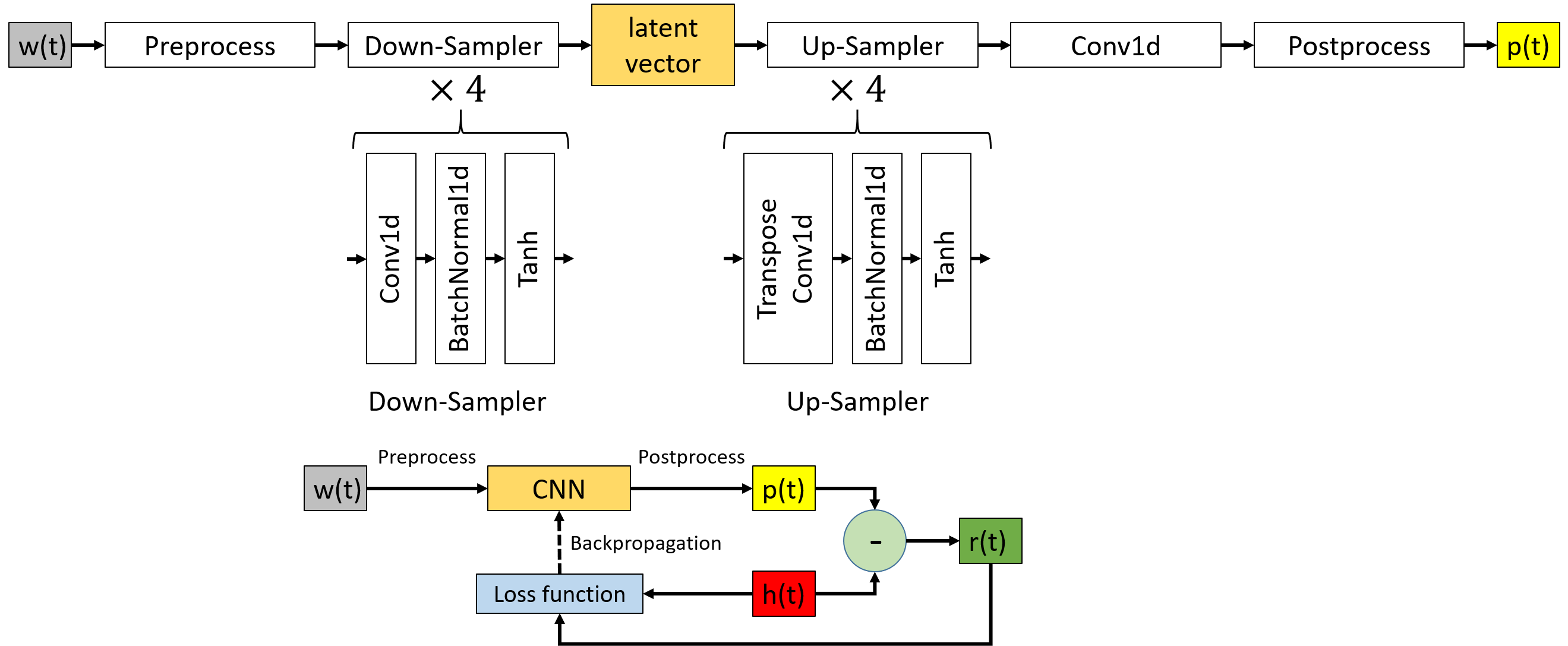}
    \caption{The top diagram illustrates the \textit{DeepClean} architecture and the workflow. \textit{DeepClean} takes timeseries data from multiple witness channels as input and runs it through a fully convolutional autoencoder. The autoencoder has four convolution layers for downsampling and four transpose-convolution layers for upsampling. After each layer, batch normalization and a tanh activation function are applied. Finally, an output convolutional layer generates the one-dimensional noise prediction. The flowchart at the bottom depicts a typical training workflow for \textit{DeepClean}. The ADAM optimizer is employed to minimize the loss function by navigating through the gradient space.}
\label{fig:workflow}
\end{figure*}

\section{The \textit{DeepClean} infrastructure}
\label{sec:infra}

The \textit{DeepClean} architecture has been described in detail in \cite{Ormiston:2020ele}. In this section, we provide a brief overview of the algorithm. The strain readout from an interferometer, $h(t)$,  can be represented as the sum of a possible astrophysical signal $s(t)$ and the detector noise $n(t)$, such that $h(t) = s(t) + n(t)$.

The goal of \textit{DeepClean} is to minimize the noise $n(t)$ to enable the detection of the astrophysical signal at the highest possible signal-to-noise ratio (SNR). While some noise sources are fundamental and cannot be eliminated, others can be removed with the help of witness sensors~\cite{2020PhRvD.102f2003B}. We can classify the noise into two categories: \textit{witnessed} and \textit{non-witnessed} noise. The environmental and instrumental processes that contribute to the witnessed noise $n_w(t)$ are monitored by a set of channels denoted as ${w_i(t)}$, as discussed in \cite{Ormiston:2020ele}. Mathematically, the noise contributed by these channels can be expressed as an output of some activation function $\mathcal{F}$, \textit{i.e.}, $n_w(t) = \mathcal{F}({ w_i(t)})$.

In general, the activation function $\mathcal{F}$ involves non-linear and non-stationary couplings, particularly in gravitational-wave interferometers. \textit{DeepClean} is a convolutional neural network that encodes this activation function using trainable weights $\vec{\theta}$. Thus, we can express the neural network as 
\begin{equation}
    n_w(t) = \mathcal{F}({ w_i(t)}; \vec{\theta}).
\end{equation}

The \textit{DeepClean} architecture was designed to be a symmetric auto-encoder that has four downsampling layers (convolution) and four upsampling layers (transpose convolution). 
The input layer has a flexible dimensionality to match the sampling frequency and number of witness channels in the input data. the first downsampling layer is designed to have 8 channels (features) with the same sampling frequency as in the input data. Each sucessive layer downsample the data by a factor 2 and increases the number of features by a factor 2, meaning that the latent vector has 64 features. 
The four up-sampling layers halves the number of features and doubles the sampling frequency at each layer, thereby regaining the same dimensionality as the input data. An output convolutional layer is then applied to map the data into a one-dimensional time series of noise prediction. At each layer, convolution or transpose convolution is followed by batch normalization and a {\it tanh} activation function to improve the model's generalization ability. A schematic diagram of the \textit{DeepClean} architecture, along with a flowchart of a typical \textit{DeepClean} workflow, is presented in Fig.~\ref{fig:workflow}. 

The weights $\vec{\theta}$ are trained using the gradient descent algorithm~\cite{2016arXiv160904747R} by minimizing an appropriate loss function. In the case of \textit{DeepClean}, the loss function is defined as the ratio of the noise spectrum of the cleaned strain to the original strain, summed over all frequency bins within the analysis bandwidth $[f_{min}, f_{max}]$:
\begin{equation}
J = \frac{1}{N} \sum_{i=1}^N \sqrt{\frac{S_r^{(i)}}{S_h^{(i)}}}
\label{eq:loss-function}
\end{equation}
Here, $S_r^{(i)}$ is the power spectral density (PSD) of the residual strain at $i^{th}$ frequency bin after subtracting $n_w(t)$. Likewise, $S_h^{(i)}$ is the PSD of the original strain at $i^{th}$ bin before subtracting $n_w(t)$.

Prior to processing with \textit{DeepClean}, both the strain and witness time-series are pre-processed by normalizing the time-series to ensure they have zero mean and unit variance. The strain data is further bandpass filtered to the frequency range of interest $[f_{min}, f_{max}]$. The pre-processed data is then input into the trained \textit{DeepClean} to predict the noise contamination. To prevent boundary artifacts, predictions are made on 8\,s segments with 4\,s overlaps. These overlapping noise predictions are then combined after applying \textit{Hann} windows to improve the prediction quality.

Subsequently, the predicted noise is band-pass filtered to $[f_{min}, f_{max}]$ to exclude any frequencies outside this range. After reversing the normalization steps, the predicted noise is subtracted from the original strain, yielding the cleaned strain.

In the following sections, we will use a mock data challenge to evaluate performance of \textit{DeepClean} and to conduct validation tests.

\section{A Mock Data Challenge}\label{sec-MDC}

To evaluate the effectiveness of \textit{DeepClean}, we performed an end-to-end analysis of mock data through a Mock Data challenge (MDC) introduced by the LVK to benchmark and prepare the low latency analysis pipelines. The mock data is generated by injecting compact binary signals into the O3 strain data from LIGO Hanford and LIGO Livingston. We selected the low-latency O3 data (labeled as \verb|GDS-CALIB_STRAIN|) from the 20-day period between September 1, 00:00:00 UTC and September 20, 00:00:00 UTC.
This period exhibits high coherence between the strain and intended witness channels 
in both H1 and L1, making it well-suited for testing the performance of \textit{DeepClean}.

The injected compact binary signals comprise binary black hole mergers, binary neutron star mergers, and neutron star-black hole mergers. The parameters of the injections such as masses, spins, luminosity distance, and other extrinsic parameters, are drawn from simulated distributions that are consistent with the O3-inferred population models~\cite{GWTC3pop} 
The coalescence times of the 25,000 injections are uniformly distributed over the 20--day period, such that there are no overlapping signals. Additionally, all the signals are generated by using 10\,Hz as the  lower cut-off frequency.

\section{Applying \textit{DeepClean} on the mock data}
\label{sec:inference}

The noise regression analysis is performed in two steps; training and cleaning (also referred to as the \textit{inference}). Below we describe the operational parameters of training and cleaning considered in this study.

\subsection{Strategy for Training and Cleaning the MDC data}

\begin{figure*}
    \centering
    \includegraphics[scale=0.36]{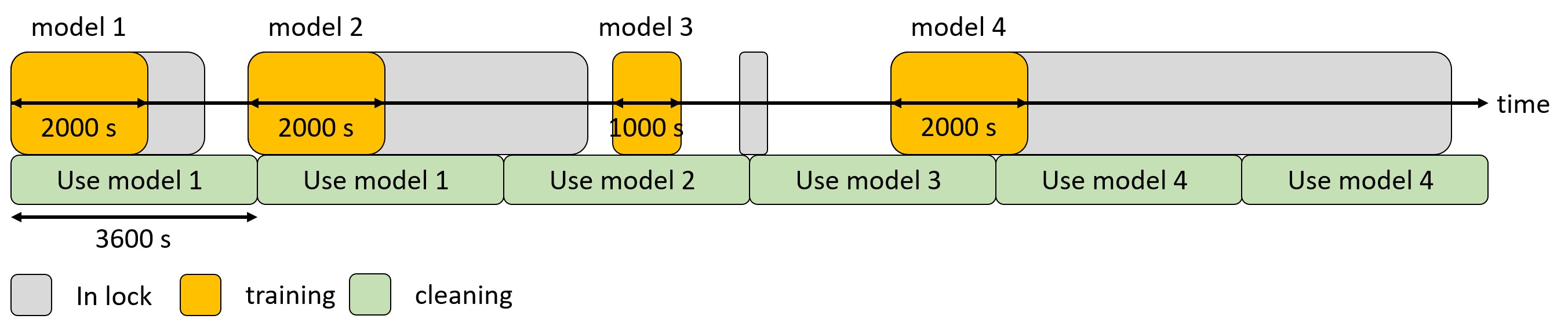}
    \caption{This schematic shows the training strategy used for analyzing the mock data. The grey shaded segments represent science-quality data, and the yellow indicates that a model training is performed at the beginning of each science segment. The green segments represents one-hour long inference periods where the trained model is used. That means, once a model is trained at the beginning of a science segment, all the subsequent data until the start of next science segment is cleaned using that model.}
\label{fig:retrain}
\end{figure*}

To train and clean the 20 days of MDC data, we adopted a strategy that involves selecting only science-quality data labeled as \verb|DMT-ANALYSIS_READY:1|. In total, we identified 47 science sub-segments (\textit{a.k.a} active segments)\footnote{For the time between two science segments, the detector is either not collecting data or the collected data does not pass data quality tests} in H1 and 72 in L1 over the 20-day period. We used \textit{DeepClean} to clean each science segment, with training performed once using the first 2000\,s of the sub-segment, regardless of the length. This approach is supported by a detailed study outlined in section~\ref{sec:training}. A visualization of this strategy can be seen in Fig.~\ref{fig:retrain}.   

\subsection{Target noise: power-line at 60\,Hz and the side-bands}

To illustrate non-linear and non-stationary couplings, we consider the 60 Hz line of the power mains, which is modulated by low-frequency noise from LIGO's alignment sensing and control (ASC) system~\cite{Tiwari:2015ofa}. This coupling produces sidebands around the central frequency, and we use a set of witness channels that were previously used to subtract these sidebands during the third observing run (O3)~\cite{Vajente_2020}.

\subsection{Data Pre-processing}

The original strain data, recorded at a rate of 16384 Hz, is downsampled to 4096 Hz for this analysis. This is because the instrumental and environmental noise we want to subtract mostly occurs at frequencies below 100\,Hz, and 4096\,Hz is sufficient for downstream analyses like detection and parameter estimation of compact binary mergers. Although \textit{DeepClean} can handle any sampling rate, a reduced rate makes training easier due to the reduced data size.

Most witness channels that are coupled to 60\,Hz power-line noise have lower sampling rates than the strain data, often below 100\,Hz or even below 10\,Hz (known as fast and slow channels). However, \textit{DeepClean} requires all input channels to have the same sampling rate. Therefore, we upsample these channels to match the strain data rate. As discussed in Sec.~\ref{sec:infra}, each channel's data is normalized independently to have zero mean and unit standard deviation. The strain data is bandpass filtered to limit the relevant frequency range for the target noise. Specifically, \textit{DeepClean} uses an 8th-order Butterworth filter to bandpass filter the data to the 55-65 Hz range. The frequency range considered here is wide enough to contain all the sidebands around 60 Hz.

\subsection{Training}

During the training process, the pre-processed data is divided into overlapping segments \textit{a.k.a} kernels\footnote{Not to be confused with the filter kernel used by the convolution operator in the CNN architecture}. These kernels are then grouped into batches, with each batch consisting of a fixed number (\texttt{batch\_size}) of kernels. For this analysis, we used a \texttt{batch\_size} of 32 with kernels that are 8\,s long and with  7\,s overlap between two kernels. This results in 25\,s of data in each batch. The entire training data of 2000\,s is then composed of 1993 overlapping kernels, which translates to a total of 80 batches. When the entire training data is passed through \textit{DeepClean} once (known as one epoch), the algorithm takes 80 iterations, with one batch taken at each iteration. At every iteration, the loss function is calculated, and backpropagated to compute the gradients, and subsequently updates the weights. In the process of  weight optimization, \textit{DeepClean} uses \textit{ADAM} optimizer\cite{AdamOptimizer} to navigate through the gradient space and minimize the loss function.

Our analysis indicates that the loss function converges in approximately 20-25 epochs during typical 60 Hz noise subtraction. This translates to roughly 1600-2000 iterations using the settings described in this example. 

\begin{figure*}
    \centering
    \includegraphics[width=0.495\textwidth]{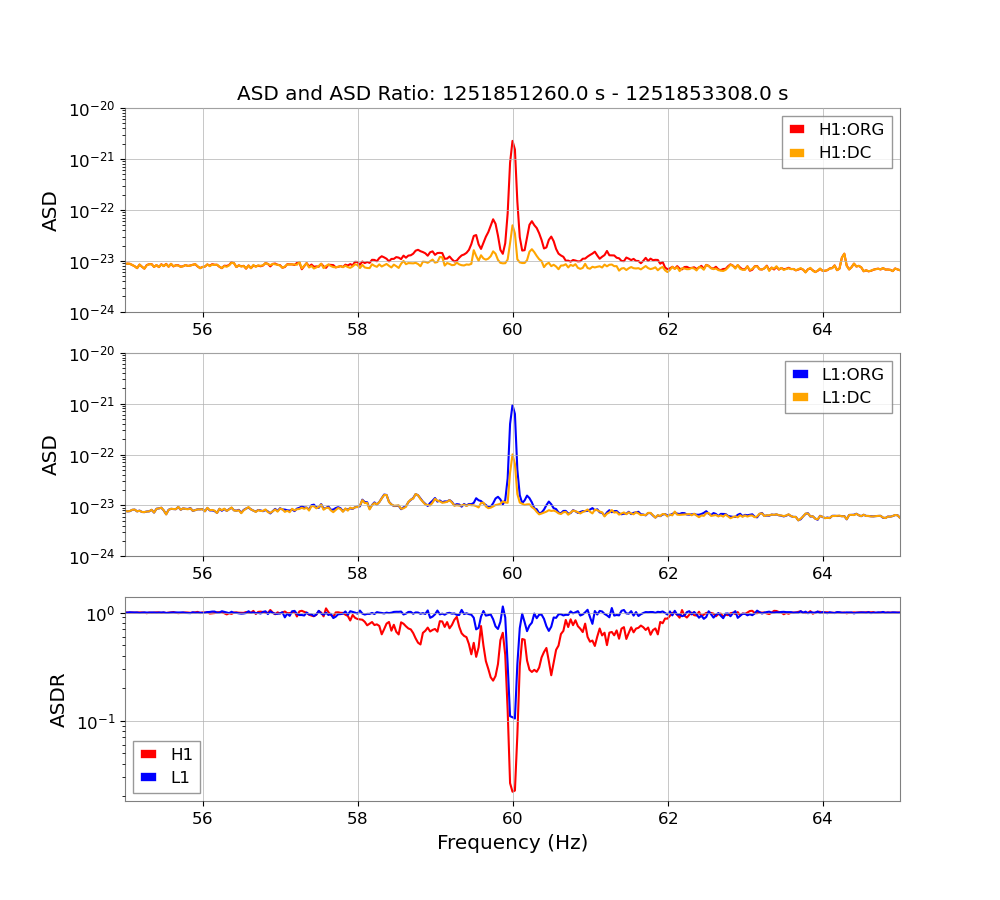}
    \includegraphics[width=0.495\textwidth]{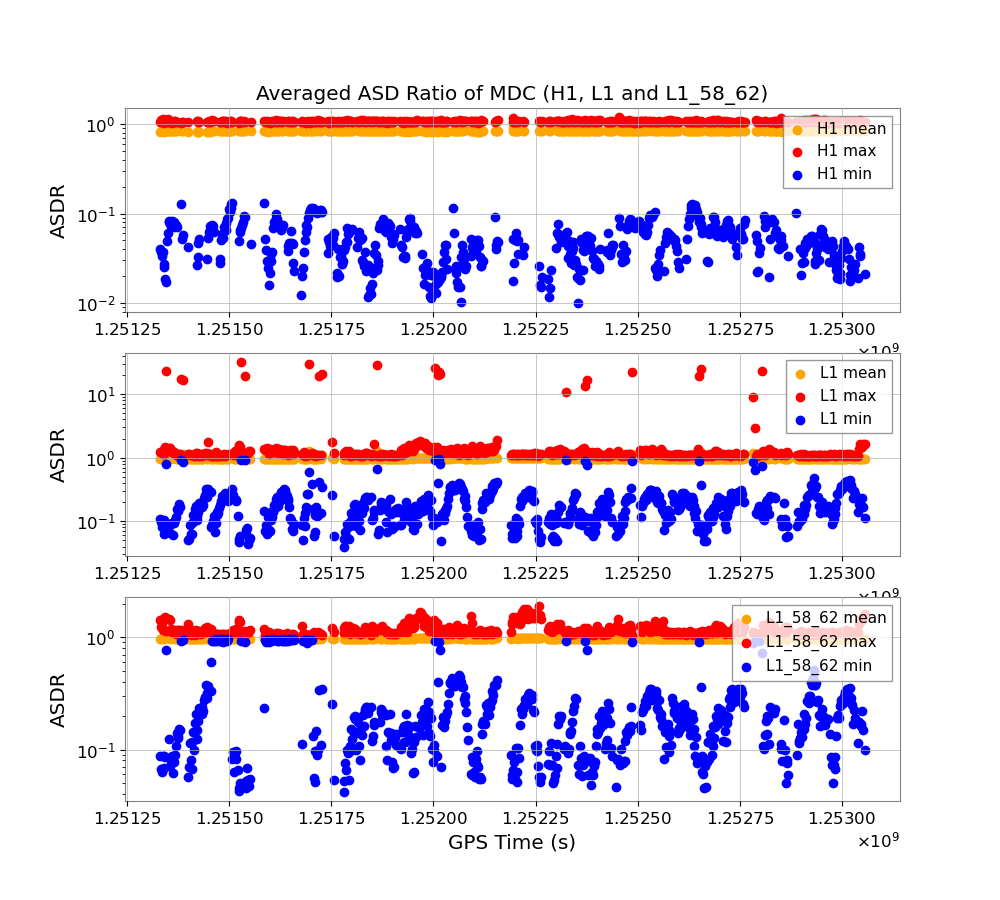}
    \caption{[Left] Amplitude spectral densities (ASD) of the original and cleaned data are shown in the top (H1) and middle (L1) panels. The bottom panel shows the cleaned-to-original ASD ratio for both H1 (red) and L1 (blue). The plots are made using randomly picked 2048 seconds of MDC data from both H1 and L1. [Right] ASD ratios computed over the 20-day period of MDC data for H1 (upper) and L1 (middle and lower). Each point on the x-axis represents 256\, s of data starting from there on and the y-axis shows the minimum (blue), maximum (red) and the mean (orange) of the ASD ratio from the [55, 65] Hz band. For L1, due to quality issues (see the descussion in the text), the analysis is repeated with a narrower frequency band;z [58, 62] Hz which is shown in the bottom panel while the middle panel shows the results obtained with [55, 65] Hz.}
    \label{fig:asdr-mdc}
\end{figure*}

\subsection{Cleaning and post-processing}\label{sec-cleaning}

Unlike in training, we do not necessarily have to perform cleaning on overlapping kernels because, given a set of trained weights, the predictions for a certain data segment are always the same and do not benefit from averaging over overlapping kernels. However, when performing bulk offline analysis of thousands of seconds of data, it is inevitable to clean the long data segment after splitting into batches of shorter segments, due to memory constraints.

In addition, CNN architecture underweights the edges of the segments by design, which can lead to artifacts at the kernel edges. These artifacts get enhanced during the bandpassing step in the post-processing, and they can also spread to samples farther from the edges. To prevent these edge effects, we apply Hann windows to the predictions from overlapping kernels and employ an averaging procedure before they are band-passed.

For the offline analysis, we perform inference on 3600\,s-long (1 hour) chunks of data, using a  model that is trained at the beginning of the science segment, as previously described.

\subsection{Performance: Improvements in the noise spectral density}

To assess the quality of noise regression, we compared the ASD of the cleaned strain to the original strain in the frequency band of 55-65 Hz using ASD ratio as a metric. The ASD ratio was computed on 2048\,s of data from both H1 and L1 data, and the resulting plot (Fig.~\ref{fig:asdr-mdc}, left panel) showed a well-subtracted peak at 60\,Hz and its sidebands.  The right panel in Fig.~\ref{fig:asdr-mdc} shows the ASD ratio computed over the 20 days of MDC data. Each point on the x-axis (in units of seconds) represents the 256\, s data starting from that second onwards. For each x value, there are three y values, which are the minimum, maximum, and the mean of the ASD ratio of that particular 256\, s. For example, at each point, there will be an ASD ratio curve similar to the bottom panel on the left. The \textit{mininum} will represent the subtraction achieved at the 60 Hz peak. The \textit{maximum} is plotted with the intention of capturing ASD ratio that goes above 1, \textit{i.e.} any noise that is contributed by \textit{DeepClean}. The \textit{mean} is meant to showcase the overall subtraction in the band including the sidebands.   

In the top right panel, we have the ASD ratios from H1 noise subtraction. the maximum stays around 1, the \textit{mean} and \textit{minimum} below 1 consistently over the 20 days. This indicates a quality subtraction.  On the other hand, for L1 subtraction in the middle right panel, we notice that the \textit{maximum} of ASD ratios are well above one for many segments. These peaks in the ASD ratios are understood to be happening at 56\, Hz and 64\, Hz while the exact reason is not well understood. It can be a data quality issue either in the strain or the witness channels, leading to poorly converged models of the neural networks. We repeated the analysis by narrowing down the  frequency band to $58-62$\, Hz such that the frequencies of noisy peaks  (56\,Hz and 64\, Hz) are excluded from the band. The results are shown in the bottom right panel. It shows that the unwanted features are filtered out by appropriately narrowing down the frequency band of the analysis.

\section{Validation tests with Astrophysical metrics}\label{sec:offline-O3}

In the preceding section, we explored the use of \textit{DeepClean} on mock data and demonstrated improvements in the ASD ratios. This section concentrates on astrophysically-motivated validation tests to ensure the effectiveness and safety of applying \textit{DeepClean} to data containing astrophysical signals. We examine two specific areas: the impact on the sensitivity of compact binary searches, demonstrating effectiveness, and the assurance of signal integrity in source parameter estimation.

\subsection{Compact binary search sensitivity}

\begin{figure}
    \centering
    \includegraphics[scale=0.6]{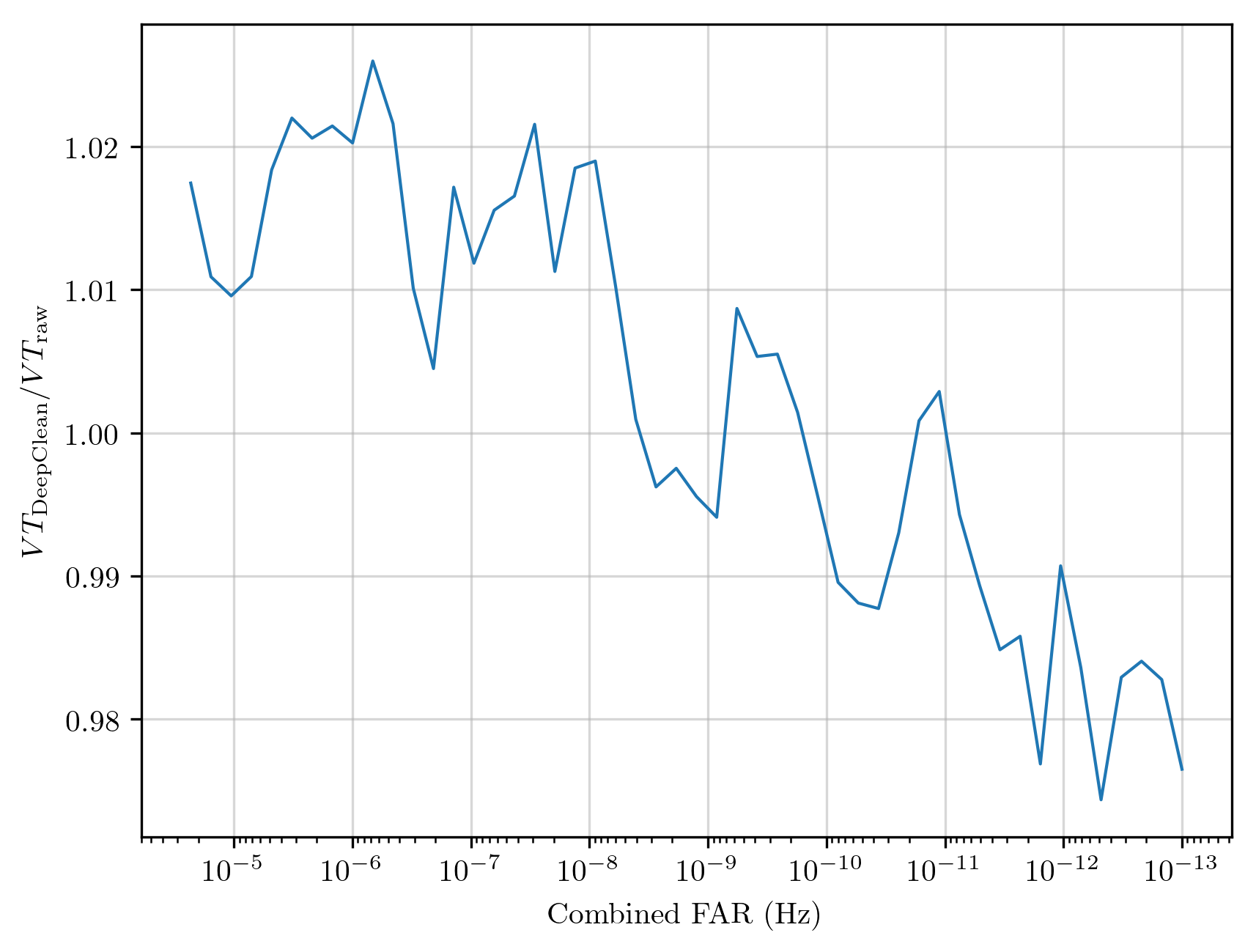}
    \caption{The fractional improvement in sensitive volume ($VT$) measured by GstLAL (after \textit{DeepClean} to before \textit{DeepClean}) shown as a function of the estimated false-alarm-rate. \textit{DeepClean} improves the sensitive volume of the search for false-alarm-rates of approximately 2 per day to 1 per 100 years, but there is a slight loss in sensitivity to very high significance events. 
    }
    \label{fig:gstlal}
\end{figure}

The GstLAL-based inspiral pipeline (referred to as GstLAL) is a matched-filtering based pipeline used to detect compact binary mergers~\citep{Sachdev:2019,Hanna:2019ezx,Messick:2017,Tsukada:2023edh,2021SoftX..1400680C}. GstLAL has played an instrumental role in low-latency detections of gravitational-waves~\citep{LIGOScientific:2016sjg,AbEA2017b}, and directly enabled the observation of electromagnetic counterparts associated with a BNS merger~\citep{LIGOScientific:2017ync,G298048}.

We perform two GstLAL analyses on $\sim$\,20 days of O3 data to assess the performance of \textit{DeepClean}. The first analysis acts as a control and uses the final strain frames cleaned and published by Advanced LIGO and Advanced Virgo~\citep{LIGOScientific:2023vdi}. The second analysis ingests the frames processed by \textit{DeepClean}. In each analysis, we search for a set of astrophysically distributed simulated gravitational-wave signals, or \emph{injections}. The injections span $m_i \in [5 M\odot, 50 M\odot]$ in component mass and $s_{i\rm{z}} \in [0, 0.99]$ in component spin aligned with the orbital angular momentum. We evenly space the resulting $54 000$ injections 32\,seconds apart, and separately place them into each data stream. 

We find that we recover more injections in data cleaned by \textit{DeepClean} for false-alarm-rates between 2 per day and 1 per 100 years, as shown in Fig.~\ref{fig:gstlal}. For highly significant simulations (false-alarm-rates less than 1 per 100 years), here is a slight loss in sensitivity. We do not expect that this loss significantly impacts the chance of detection for these loud events. We hypothesize that this behavior is a result of \textit{DeepClean} focusing on removing quiet noise artifacts while leaving loud noise transients from other sources in the data, causing the slope of the extrapolated background to lessen. In the limit of more data, we expect the $VT$ ratio at high significance to asymptote to 1; we leave confirmation of this behavior to future work.

\subsection{Parameter estimation of coalescing binaries}

After applying a denoising pipeline, it is critical to perform parameter estimation (PE) of the underlying astrophysical signals as a validation test. This serves two purposes: firstly, to ensure that the regression analysis has not affected the original signals and, secondly, to assess any improvement in the credible intervals of the estimated parameters resulting from noise-subtraction. In this study, we focus only on the first purpose since the noise subtraction of 60 Hz alone may not yield any notable improvement in the credible interval. 

To perform this test, we selected injections from our MDC dataset based on their coalescence frequencies. The coalescence frequency, also known as the frequency of the last stable circular orbit of a binary evolution, is given by $f_{lso} = (6^{3/2}\,\pi\, m)^{-1}$ Hz, where $m=m_1+m_2$ is the total mass of the binary in the observer frame. We pick only those injections whose $f_{lso}$ lies between 55\,Hz and 70\,Hz. We choose this frequency range because our target noise is around 60\,Hz  and signals with a peak frequency around 60\,Hz would demonstrate the most significant scientific benefits. We found 258 BBH injections that satisfy this source criterion.

The literature contains well-described methods for estimating parameters from gravitational-wave signals, and there are standard analysis pipelines available that use stochastic samplers \citep{Veitch:2014wba,Ashton:2018jfp}. For our analysis, we utilized tools from the \textit{Bilby} \citep{Ashton:2018jfp} Bayesian library. We ran the \textit{Dynesty} sampler~\cite{2019S&C....29..891H} to sample from a 15-dimensional parameter space that included the luminosity distance, two mass parameters, six spin parameters, the time and phase of the binary coalescence, and four angles defining the binary's sky-location and orientation relative to the line of sight.  

Out of the 258 injections, only 84 events met the minimum signal-to-noise ratio criterion of 4 at both detectors, confirming their detection and indicating the potential for reliable parameter estimation. Additionally, we did encounter sporadic instances where the cleaned data was noisier than the original data. We subsequently excluded these affected segments from our analysis and we were then left with 78 injections for our PE study. In order to address this issue for practical online setups, we need to incorporate validation tests to ensure that the outputs of the \textit{DeepClean} algorithm are not noisier than the original. If the \textit{DeepClean} output is found to be noisier, one needs to replace them with the original data as a baseline solution. More involved approaches to resolve this issue would include increasing the cadence of training.

\begin{figure}
    \centering
    \includegraphics[scale=0.4]{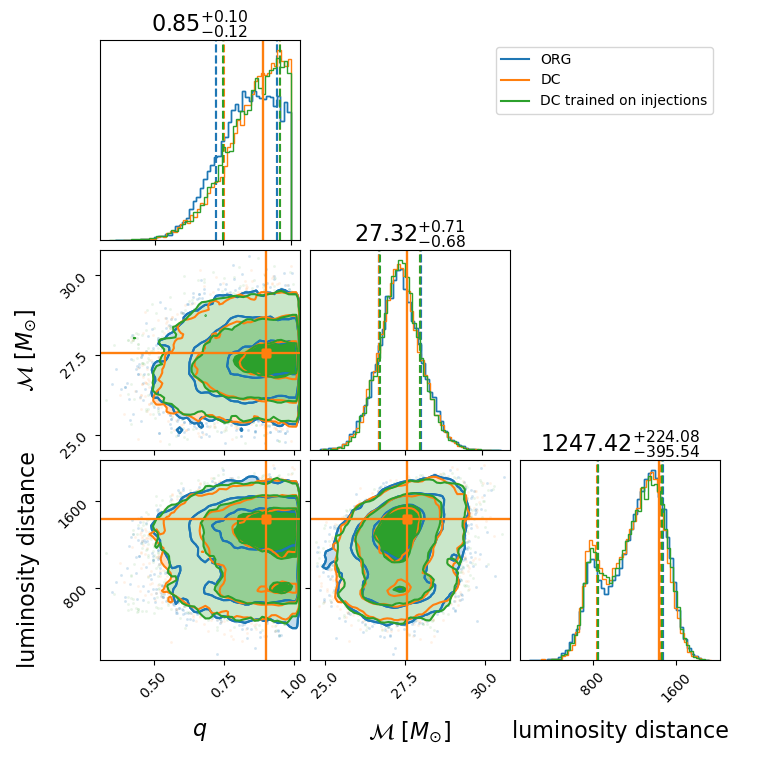}
    \caption{ 
    Corner plot showing the posteriors of the mass parameters and the luminosity distance before and after the subtraction of the 60 Hz power-line and their side-bands using \textit{DeepClean}}
    \label{fig:corner-3param}
\end{figure}

We conducted parameter estimation (PE) on both the denoised and original strain data, and compared the results. 
In Fig.~\ref{fig:corner-3param}, we present 3D posteriors of the luminosity distance and two mass parameters obtained from one of the 258 injections we analyzed. The posteriors from the cleaned data (orange) are consistent with those of the original data (blue). This indicates that the noise regression analysis did not introduce any unwanted noise components or remove any spectral features from the signal itself. The same is true even if \textit{DeepClean} is trained on data that has injections, as shown in the green curve.   

\begin{figure*}
    \centering
    \includegraphics[scale=0.53]{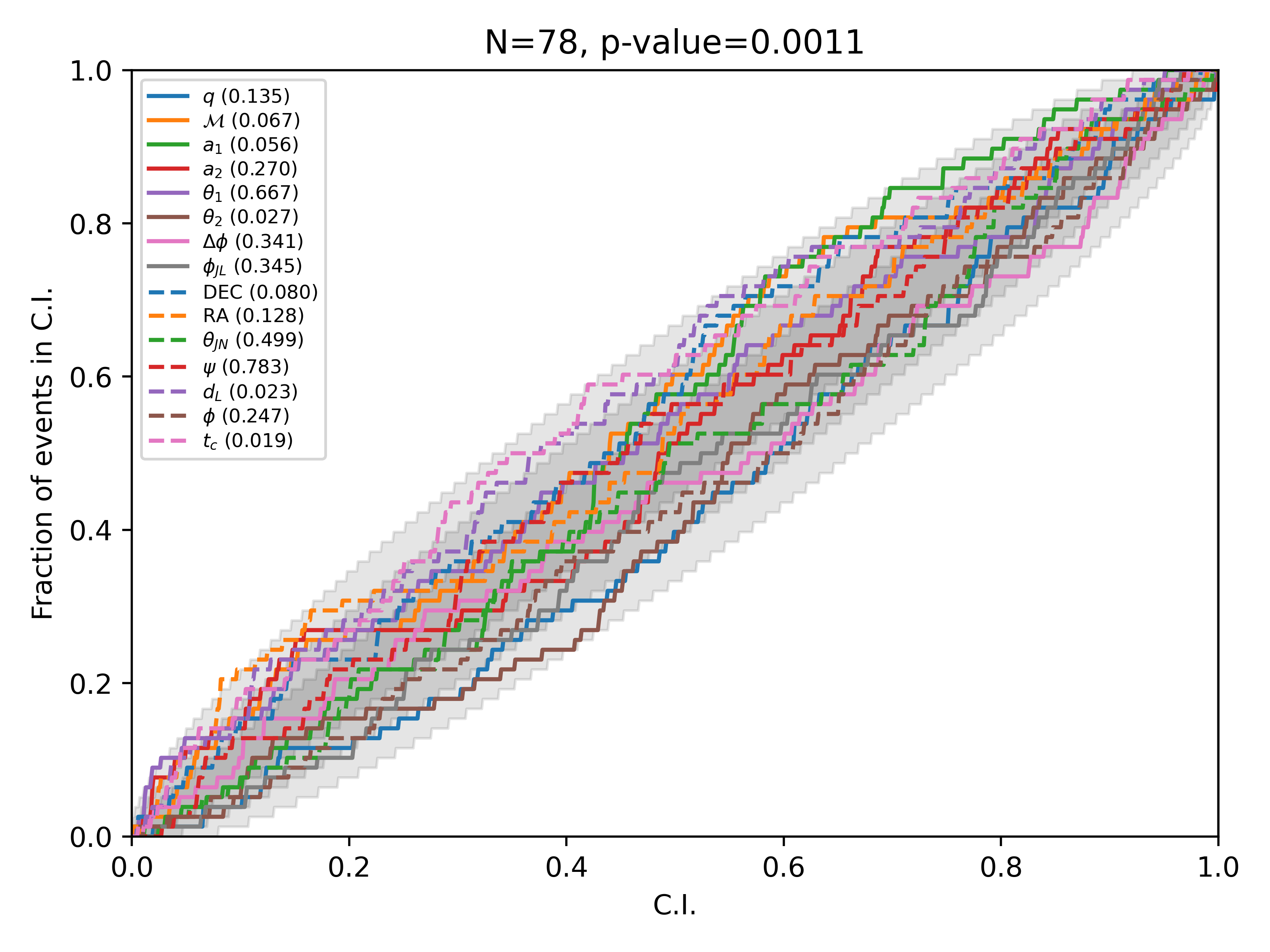}
    \includegraphics[scale=0.53]{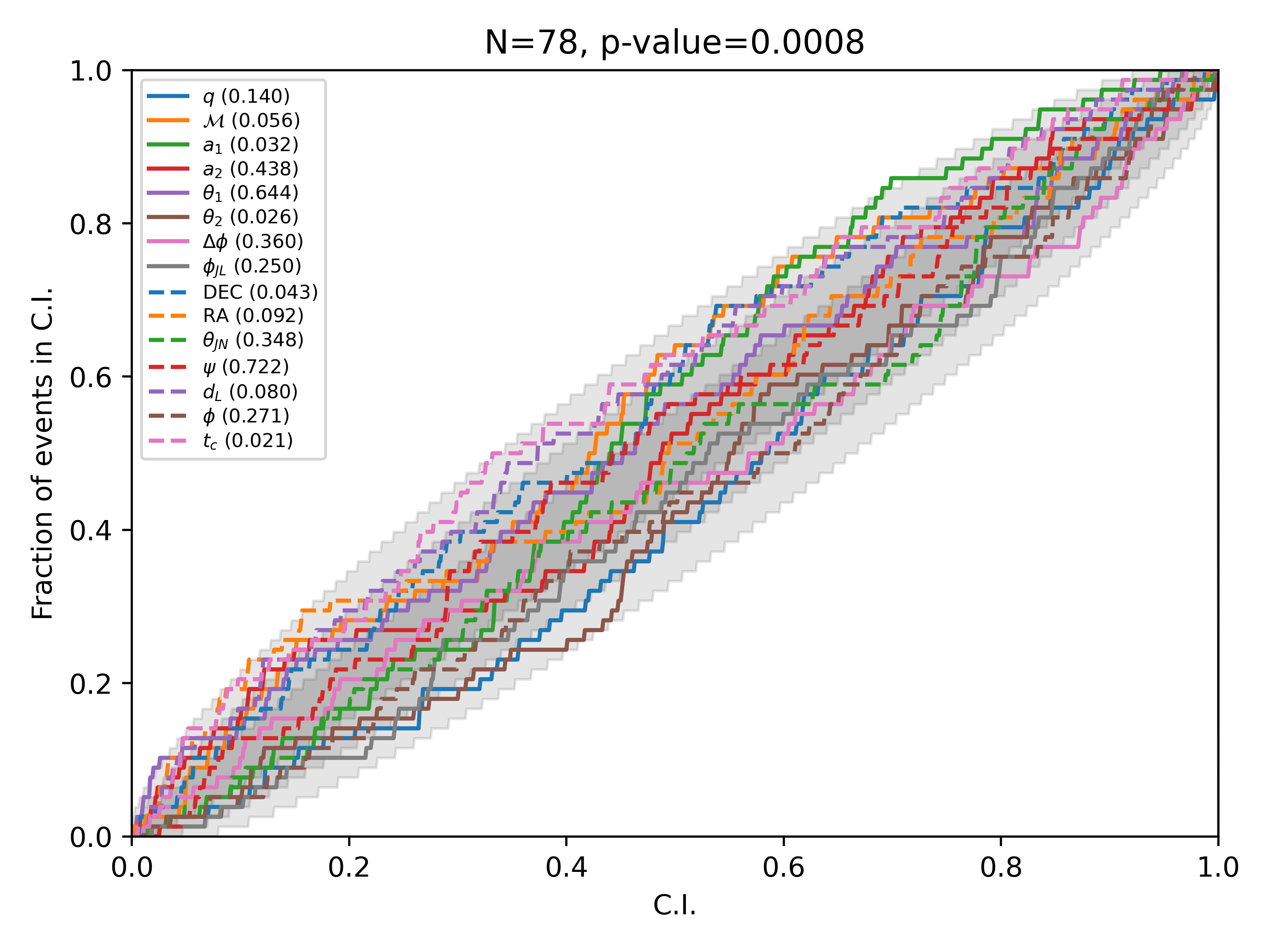}

    \caption{P-P plots generated from the parameter estimation studies of 78 binary black hole injections, comparing the results before and after the application of the \textit{DeepClean} algorithm for offline noise subtraction. The x-axis represents the credible interval, while the y-axis shows the fraction of injections recovered within that interval. These P-P plots are used to validate whether the injected parameters, after noise regression, can be recovered within the statistical uncertainty limits. As seen in the figure, the parameter recovery after \textit{DeepClean} is at least as good as, if not better than, that achieved prior to applying the algorithm. The P-values included in the plots are derived from the Kolmogorov-Smirnov test, which quantifies the degree to which the credible interval distributions differ from the expected distributions.}
    \label{fig:pp-plots-offline}
\end{figure*}

Fig.~\ref{fig:pp-plots-offline} displays p-p plots for the fifteen parameters from the 78 events, showing excellent agreement between the p-p plots before \textit{DeepClean} (left) and after \textit{DeepClean} (right). This observation is essential as it validates the safety of the underlying astrophysical signals when the \textit{DeepClean} algorithm is applied. This result demonstrates that the algorithm does not harm the underlying astrophysical signals and hence supports the reliability of the analysis.

\section{Feasibility Study for Low-Latency Deployment of \textit{DeepClean} }\label{sec:dc-online}

To perform noise subtraction in real-time, we must employ a different approach from the offline analysis outlined in the previous section. The offline analysis involved dividing the data into chunks of 3600\,s, predicting noise on overlapping kernels of 8\,s, and then combining them after applying window functions (see section~\ref{sec:inference} for more details). This approach necessitates having a substantial amount of data available at once, enabling the creation of overlapping 8\,s kernels. In contrast, the online analysis aims to clean the data as soon as it becomes available and make it accessible to low-latency search pipelines downstream of \textit{DeepClean}. Therefore, we need a different workflow and strategy for the online version of \textit{DeepClean}. A fully functional model of online \textit{DeepClean} complemented with the \textit{Inference-as-a-service} model will be presented in a future publication. Here, we discuss the key differences that separate it from the offline model, the issues it raises, and some preliminary figures of merit.

\subsection{Edge effects}

\begin{figure*}
    \centering
    \includegraphics[scale=0.86]{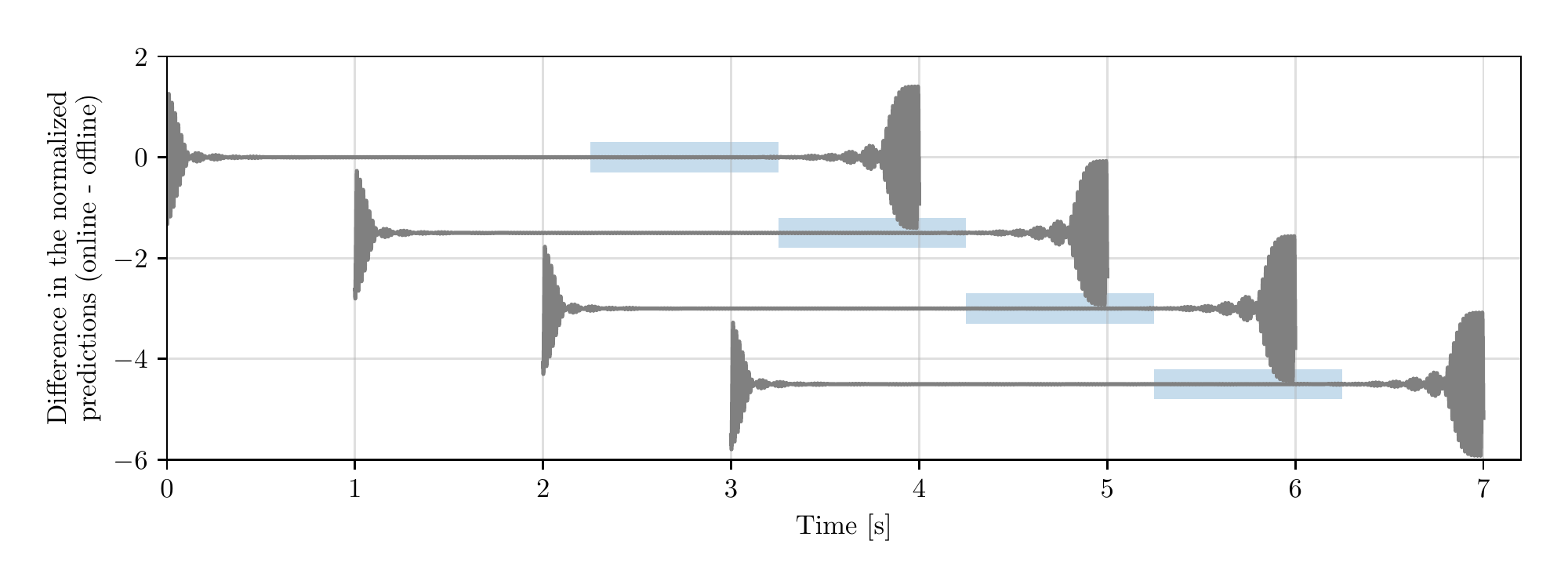}
    \caption{The diagram shows how \textit{DeepClean} will be used for low-latency (online) denoising. The figure compares the online and offline predictions, both normalized, for 4-second duration segments. The offline prediction for the same segment is extracted from the middle of a 128-second-long prediction. The difference bars indicate that the predictions differ only at the edges, which are less than a second long. To avoid edge effects, \textit{DeepClean} is applied to 8 seconds of data, consisting of 6 seconds from the past and 1 second from the future, in addition to the 1 second of target data. This is done for every 1 second frame, and the wait for a future frame causes an additional 1 second of latency. After the prediction, everything except the 1 second target data is discarded from the 8-second segment and written to disk as cleaned frames, ready for downstream analyses.}
\label{fig:edge-effects}
\end{figure*}

The need to develop a new strategy arises from \textit{edge effects}, which occur when the noise prediction quality deteriorates towards the edges of a kernel. Fig.~\ref{fig:edge-effects} shows that the first and last approximately 0.5\,s of a 4\,s kernel are susceptible to noisy spectral features. The width of each 4\,s segment along the vertical axis shows the difference between the online and offline predictions. The offline prediction is made on a 128\, second longer segment in which the 4 seconds shown are far from the edges. 
To mitigate these effects, \textit{DeepClean} uses overlapping  kernels and Hann windows to give more weight to the center of each kernel. This approach has been found to work well for offline cleaning. For online cleaning, our aim is always to clean the 1\,s segment that is recorded most recently. Even if we divide the 1\, s data into shorter overlapping kernels, there are no data available to overlap with the very last kernel and hence the edges can not be fully suppressed.   

The reduction in quality at the edges of the kernel can be attributed to the natural tendency of CNN architectures to discard information at the edges of input data. This is particularly relevant in \textit{DeepClean}, where we employ filter kernels of size 7 and strides of 2, resulting in the features at the edges being captured at a lower level compared to those at the middle of the kernel during convolution. For example, a sample from the middle is captured four times by the sliding filter, while a sample at the edge is only captured once. While \textit{DeepClean} attempts to alleviate this issue by using zero-padding of size 3 at the edges, the edges are still  captured only three times with this padding size. Increasing the padding size could be a potential solution, but it could lead to array size problems since the padding size is also constrained by the input-output sample size matching.

\subsection{A working model for $\sim 1s$ latency}

While ongoing work aims to comprehensively address edge effects, a simple modification to the workflow can mitigate the issue in the meantime. This resolution comes at the cost of a latency of approximately 1 second. To ensure that the quality of our online analysis matches that of our offline analysis, we employ a 4-second kernel that includes 2 seconds of data before and 1 second after the 1-second target segment. The additional data ensures that the target segment is located in the center of the kernel. The DeepClean is then applied on the 4-second kernel, and the 1-second target segment is extracted for analysis. A cartoon depicting this is shown in Fig.~\ref{fig:edge-effects}.

\begin{figure}
    \centering
    \includegraphics[scale=0.25]{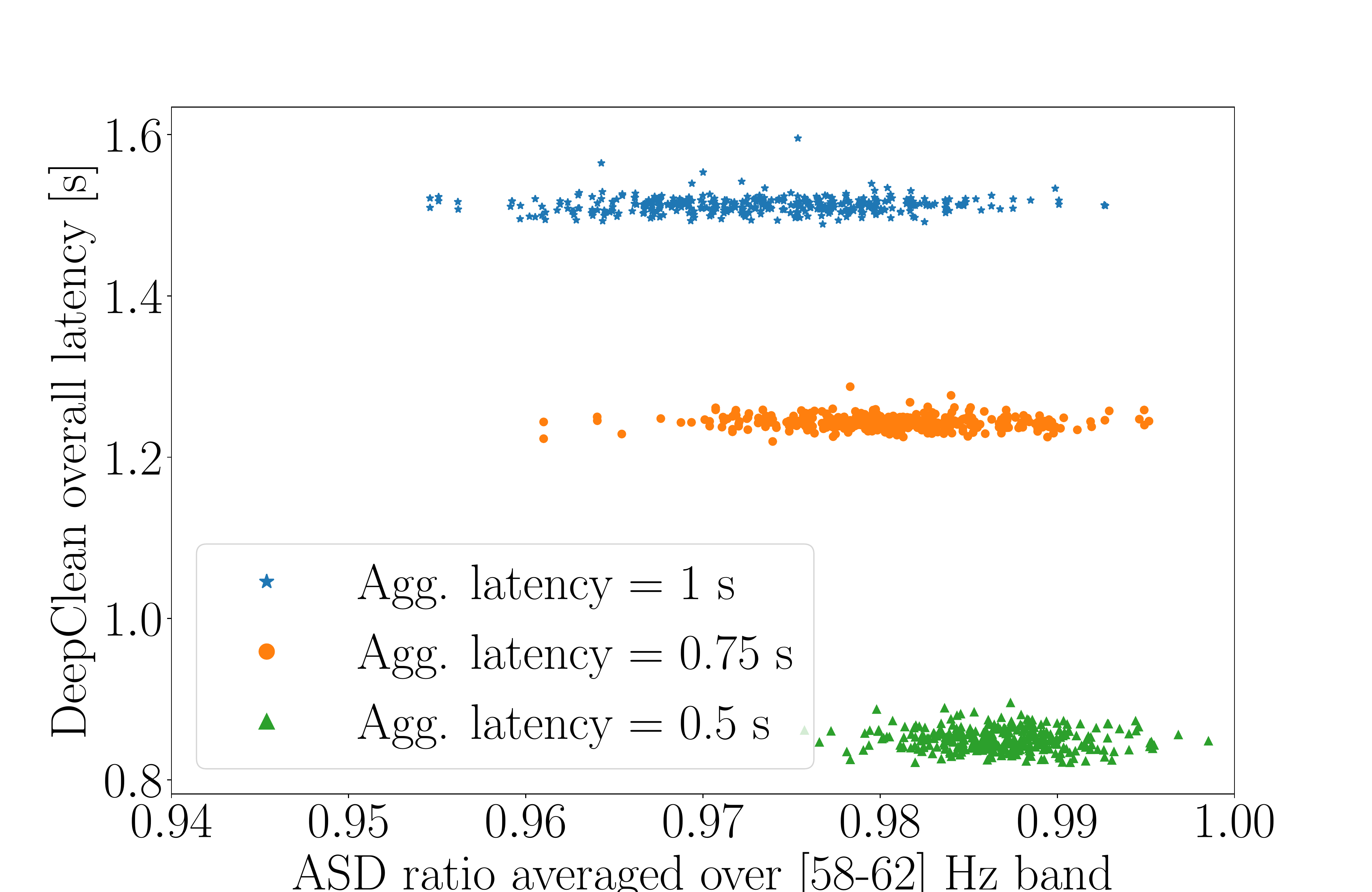}
    \caption{Scatter plot showing the latency vs quality trade-off.  The overall latency referred to here is the time taken by DeepClean to produce the output strain after the raw strain is made available. The ASD (amplitude spectral density) ratio on the x-axis is computed from every 32\,s of the data. The three different colors shows the different aggregation latencies. Notably, the quality of the ASD ratio improves with higher aggregation latency, at the cost of increased overall latency.}
    \label{fig:asdr-latency-online}
    
\end{figure}

As the affected edges are not exactly 1\,s in length, we can select a part of the output that is closer to the edge where we aggregated the future data. This edge, where future data is aggregated, is referred to as the aggregation latency. Fig.~\ref{fig:asdr-latency-online} displays a scatter plot with the achieved ASD ratio on the x-axis and the overall latency\footnote{The \textit{overall latency} is the term referred to as the latency in producing the cleaned frames \textit{w.r.t} the time when the original frame became available. It is comprised of the time taken to load the data, perform subtraction, and write the output frame to disc, in addition to the aggregation latency.} on the y-axis, for different aggregation latencies. It is evident that the subtraction quality gets better by allowing higher aggregation latencies as shown by the reduced ASD ratios in the graph.

\subsection{A case study for training DeepClean at low latency}
\label{sec:training}

\begin{figure}
    \centering
    \includegraphics[scale=0.6]{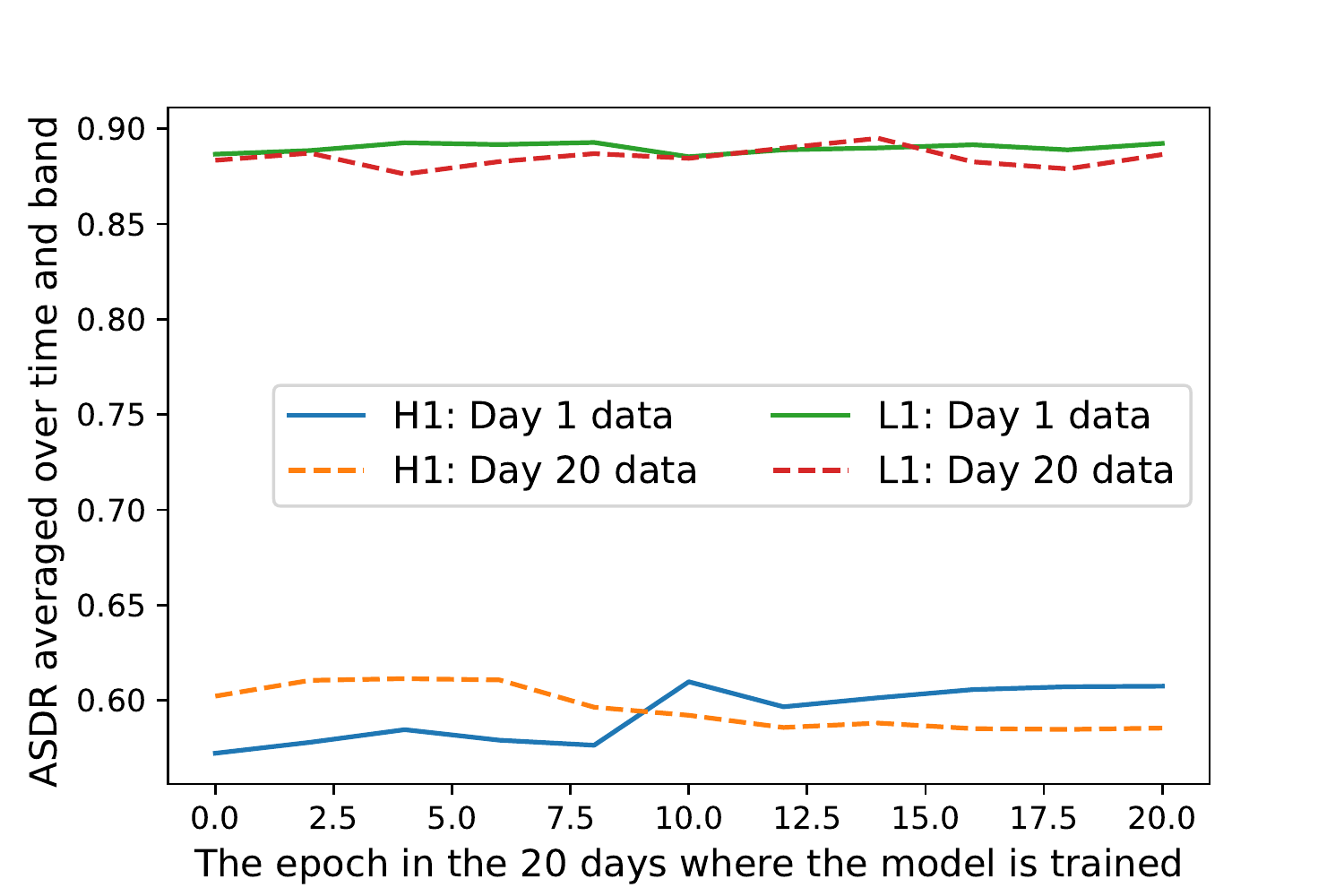}
    \caption{The figure underscores the significance of periodically retraining the neural network, \textit{deepClean}. It presents two traces for each detector: one representing the cleaned strain data from day 1 (solid line) and the other displaying the cleaned strain from day 20 (dashed line), derived from our 20-day MDC dataset. The horizontal axis depicts the GPS time of the training data, with an initial time conveniently set to zero. Meanwhile, the vertical axis represents the Amplitude Spectral Density Ratio (ASDR). Notably, there is an observed increase in ASDR when the training data is selected from a segment further away from the cleaning data, particularly noticeable in H1. In contrast, the disparity in ASDR between training and cleaning data is less prominent in L1 over this 20-day period. Overall, the figure emphasizes the importance of regular retraining to ensure optimal performance and accurate data cleaning.}
    \label{fig:training-vs-asdr-hist}
    
\end{figure}

To enable real-time gravitational wave data cleaning, it is necessary to train and validate the machine learning model in short timescales. Unlike offline analysis where there is sufficient time to optimize and fine-tune the trained model, online cleaning requires that the model be trained quickly and validated frequently. This is because the noise features in the data are generally non-stationary, and the noise coupling that \textit{DeepClean} once learned could change after a certain time, making it necessary to have new models periodically trained on the most recent data.

To explore this in detail, we conducted a case study using the 20 days of MDC data. We trained the model once for each science segment, resulting in a total of 47 trained models for H1 and 72 for L1 over the 20 days. We then took two examples of inference data, one from day 1 and the other from day 20, and cleaned them with all the models available. Fig.~\ref{fig:training-vs-asdr-hist} shows the ASD ratio results on the y-axis and the time (from the 20 days) where the model is trained on the x-axis. The solid line represents the data from day 1, and the dashed line represents the data from day 20, for both the detectors. 

It is observed that the ASD ratio changes as we move away from the time of the training dataset, particularly in H1 data. For instance, the day 1 data has an ASD ratio below 0.6 when trained with data from day 1, 2, or 3, but the ASD ratio goes above 0.6 when the training dataset is from day 20. The same observation is true for cleaning the data from day 20, which is best cleaned with the model trained on day 20.

Although this study indicates that trained models become sub-optimal over time, we noticed that it does not happen over the timescale of minutes or hours, but rather changes over the timescale of days. In L1, we do not notice any significant change in the ASD ratio over time, which may be an indication that the coupling features in L1 are rather stationary. Overall, our observations emphasize that there is enough time for training and validating a model that is trained at low latency.

It should be noted that these observations are based on our cleaning of 60\,Hz powerline and sidebands. The timescale over which the coupling features change will also depend on the coupling itself. For example, a different coupling in a different frequency range, can be highly non-stationary and would require model retraining at a higher cadence. Fortunately, our production deployment of \textit{DeepClean} is highly compute-efficient due to leveraging the GPU resources and hence capable of training once every 30 minutes or less. The details of that is differed for a later publication.

\section{Summary and Outlook}\label{sec:conclusions}

We conducted noise regression using the \textit{DeepClean} algorithm on bulk offline O3 data with high latency. For offline analyses, we focused on optimizing the network configuration, without being constrained by latency, which is crucial for online analyses. The cleaned data obtained from offline analyses are validated using downstream applications, such as detection and parameter estimation of detected compact binaries.

In a separate analysis, we demonstrated the applicability of \textit{DeepClean} for low-latency noise subtraction, where unclean data is fed as 1-second-long frames. However, we observed some discrepancies at the edges of the cleaned segments due to the inherent nature of CNN-like architectures, which require further efforts to mitigate. To overcome this issue, we proposed a workaround by waiting for a future 1-second data, such that there are no edges in the current segment. By adopting this strategy, we were able to replicate the results obtained from the high-latency analysis, showing the effectiveness of the low-latency \textit{DeepClean} application.

To evaluate the efficacy of \textit{DeepClean} over time, we investigated how frequently the trained models need to be updated. Our analysis of 20 days of MDC data revealed that retraining the model every 1 or 2 days is sufficient for subtracting the 60 Hz noise. However, we acknowledge that this interval may vary depending on the nature of the coupling. The \textit{DeepClean} deployment described in Ref.~\cite{FlexScience} allows frequent model training, as often as every 30\,min, which we anticipate would be sufficient for most of the couplings we encounter in the future. 

Our ongoing work includes extending \textit{DeepClean} to different frequency ranges especially aiming the broadband noise in LIGO detectors below 30\, Hz. Efforts are also going on to apply \textit{DeepClean} on Virgo and KAGRA data. Further, as mentioned before, an end-to-end model of \textit{online-DeepClean} is being built, deployed, and tested with different validation methods. This comes as part of preparing for production application of \textit{DeepClean} in O4. The details pertaining to all the ongoing efforts will be detailed in future publications.

\section*{Acknowledgements}
M.S., W.B., and M.C. acknowledge the support from the National Science Foundation with grant numbers PHY-2010970 and OAC-2117997.
D.C. acknowledges support from NSF Grants No. OAC-2117997 and No. PHY-1764464.

Thanks are due to computational support provided by LIGO Laboratory and supported by National Science Foundation Grants PHY-0757058, PHY-0823459. This material is based upon work supported by the NSF's LIGO Laboratory which is a major facility fully funded by the National Science Foundation. This research has made use of data obtained from the Gravitational Wave Open Science Center (www.gw-openscience.org), a service of LIGO Laboratory, the LIGO Scientific Collaboration and the Virgo Collaboration. 
Virgo is funded by the French Centre National de Recherche Scientifique (CNRS), the Italian Istituto Nazionale della Fisica Nucleare (INFN) and the Dutch Nikhef, with contributions by Polish and Hungarian institutes.
This work makes use of  \textsc{NumPy} \cite{vanderWalt:2011bqk}, \textsc{SciPy} \cite{Virtanen:2019joe}, \textsc{Matplotlib} \cite{Hunter:2007}, \textsc{jupyter} \cite{jupyter}, \textsc{corner} \cite{corner} software packages. 
We thank Siddharth Soni for his useful comments on the manuscript. This paper has been assigned the internal LIGO preprint number {P2300153}.

\bibliography{references.bib} 
\bibliographystyle{apsrev}

\end{document}